\documentclass[aps,superscriptaddress,showpacs,nofootinbib,twocolumn]{revtex4}
\usepackage{bm}
\usepackage{xcolor}
\usepackage{graphicx,color}
\usepackage{amsmath}
\usepackage{amssymb}

\newcommand{\e}{\epsilon}

\newcommand{\ra}{\rightarrow}

\newcommand{\sg}{\sigma}

\newcommand{\pa}{\partial}

\newcommand{\na}{\nabla}

\begin{document}

\title{Superfast amplification and superfast nonlinear saturation of perturbations as the mechanism of turbulence}

\author{Y. Charles Li}
\email{liyan@missouri.edu}
\affiliation{Department of Mathematics, University of Missouri, 
Columbia, MO 65211, USA}

\author{Richard D. J. G. Ho}
\affiliation{School of Physics,
University of Edinburgh, Edinburgh EH9 3JZ, United Kingdom}

\author{Arjun Berera}
\affiliation{School of Physics,
University of Edinburgh, Edinburgh EH9 3JZ, United Kingdom}

\author{Z. C. Feng}
\email{fengf@missouri.edu}
\affiliation{Department of Mechanical and Aerospace Engineering, 
University of Missouri, Columbia, MO 65211, USA}

\begin{abstract}
Ruelle predicted that the maximal amplification of perturbations in homogeneous isotropic turbulence is exponential $e^{\sigma \sqrt{Re} t}$ (where $\sigma \sqrt{Re}$ is the maximal Liapunov exponent). In our earlier works, we predicted that the maximal amplification of perturbations in fully developed turbulence is faster than exponential $e^{\sigma \sqrt{Re} \sqrt{t} +\sigma_1 t}$. That is, we predicted superfast initial amplification of perturbations. Built upon our earlier numerical verification of our prediction, here we conduct a large numerical verification with resolution up
to $2048^3$ and Reynolds number up to $6210$. Our direct numerical simulation here confirms our
analytical prediction. Our numerical simulation also demonstrates that such superfast amplification of perturbations leads to superfast nonlinear saturation. We conclude that such superfast amplification
and superfast nonlinear saturation of ever existing perturbations serve as the mechanism for the generation, development and persistence of fully developed turbulence.
\end{abstract}


\date{\today}

\maketitle

Turbulence is ubiquitous in nature, and most of air and water flows are turbulent.  Unfortunately such a common problem is the most 
important unsolved problem of classical physics according to Richard Feyman. From weather forecasting in meterology to airplane flight, turbulence poses the major scientific challenge. Turbulence as an open problem has two aspects: turbulence engineering and turbulence physics \cite{Li13}. Turbulence engineering deals with how to describe turbulence in engineering,
and  turbulence physics deals with the physical mechanism of turbulence. In pursuit of the understanding of turbulence physics, the 
classical hydrodynamic instability theory was developed \cite{Li18}. The scope of hydrodynamic instability thery is very limited, 
and the limited result is far from satisfactory in answering the basic question of transition from laminar flow to turbulence. With 
the development of chaos theory, the idea of chaos was introduced for undertanding turbulence \cite{RT71}. On the other hand, there 
are fundamental differences between chaos and turbulence according to experimental experts on turbulence.

Ever since Ruelle and Takens introduced the idea of chaos for understanding the mechanism of
turbulence \cite{RT71}, there has been constant effort in validating the idea \cite{BJPV05}. At moderate Reynolds number, some features of chaos are indeed embedded in transient turbulence
\cite{KK01}-\cite{LK15}. A key measure of chaos is the Liapunov exponent. Ruelle predicted that the
maximal Liapunov exponent in homogeneous isotropic turbulence is proportional to the square root of the Reynolds number $\sigma \sqrt{Re}$ \cite{Rue79}-\cite{BH18}. Extensive numerical
studies on Ruelle's prediction
have been conducted over the years \cite{CJVP93}-\cite{BH18}. Ruelle's prediction is based on the
assumption that the maximal Liapunov exponent is the inverse of the Kolmogorov time
scale which makes numerical
resolution a challenging problem. On the other hand, experimental experts on turbulence do not believe
that fully developed turbulence behaves as low-dimensional chaos. Fully developed turbulence is more violent than chaos. Chaos is often in the form of a strange attractor which is characterized by exponential amplifications of perturbations within the attractor, with the coefficient of the exponent being the so-called Liapunov exponent. We predicted that the maximal
amplification of perturbations in fully developed turbulence is faster than exponential \cite{Li14}-\cite{Li18},
\begin{equation}
e^{\sigma \sqrt{Re} \sqrt{t} +\sigma_1 t} ,
\label{OP}
\end{equation}
where $\sg$ and $\sg_1$ are two positive constants, $\sigma_1 = \sqrt{\frac{e}{2}} \sg$. When the time $t$ is small, $\sqrt{t}$ is much
bigger than $t$. Thus we predicted that initial stage maximal amplification of perturbations is
superfast (much faster than exponential). Together with the large Reynolds number in the fully developed turbulence regime, such superfast amplification of perturbations will quickly reach
nonlinear saturation during which time the second exponent term $\sigma_1 t$ (corresponding to a exponential factor) is much smaller than the first term $\sigma \sqrt{Re} \sqrt{t}$. Ruelle's prediction is based on dimension analysis of the Kolmogorov time scale \cite{Rue79}. Our prediction is based on rigorous analysis on the Navier-Stokes equations
\cite{Li14}-\cite{IL18}. We have conducted an extensive low resolution numerical verification on our
prediction \cite{FL18}. Our low resolution numerical simulations were not able to get into fully
developed turbulence regime. On the other hand, the numerical investigation showed that
our analytical prediction applies to a wide regime beyond fully developed turbulence. 
Built upon our earlier numerical verification \cite{FL18}, here we conduct a large numerical verification with resolution up to $2048^3$ and Reynolds number up to $6210$. We carry out direct numerical simulations on the three dimensional Navier-Stokes equations in the homogeneous isotropic turbulence regime. It is in such a fully developed homogeneous isotropic turbulence regime that our analytical prediction is most likely to be accurate. Our numerical results here indeed confirm our
analytical prediction. Our numerical simulation also demonstrates that such superfast amplification of perturbations leads to superfast nonlinear saturation. We conclude that such superfast amplification
and superfast nonlinear saturation of ever existing perturbations serve as the mechanism for the generation, development and persistence of fully developed turbulence.

Unlike the Lagrangian approach which tracks the trajectories of individual fluid particles, here we
use the Eulerian approach which tracks the evolution of the entire fluid field starting from an
initial condition of the fluid field. We will add a perturbation to the initial condition and track
the evolution of the perturbation either nonlinearly or linearly. By subtracting the two fluid
fields, we are tracking the evolution of the perturbation nonlinearly. By subtracting the
Navier-Stokes equations along the two fluid fields, we get the governing equations of the nonlinear
evolution of the perturbation. By dropping the nonlinear terms of the perturbation, we get the governing equations of the linear evolution of the perturbation, and by solving which we track
the evolution of the perturbation linearly.

We conduct direct numerical simulations on the three dimensional Navier-Stokes equations under
periodic boundary condition of period $(2\pi )^3$ with a de-aliased pseudo-spectral code \cite{Yof12}
\begin{equation}
\pa_t {\bf u} + {\bf u} \cdot \na {\bf u} = -\na p + \nu \Delta {\bf u} +{\bf f} , \  \na \cdot {\bf u} = 0,
\label{NS}
\end{equation}
where ${\bf u}$ is the velocity field, $p$ is the pressure, $\nu$ is the viscosity, and ${\bf f}$ is the external forcing. The external forcing will only force low wave numbers (large scales), and the
Fourier transform $\tilde{{\bf f}}$ of ${\bf f}$ is given by \cite{Mac97}
\begin{equation}
\tilde{{\bf f}} ({\bf k} , t) = \left \{ \begin{array}{l} \frac{\e_f}{2E_f} \tilde{{\bf u}} ({\bf k}, t),
\ \text{if } 0< |{\bf k} | < k_f , \cr  0, \quad \text{otherwise} , \cr
\end{array} \right .
\label{EF}
\end{equation}
where $\tilde{{\bf u}} ({\bf k}, t)$ is the Fourier transform of the velocity field ${\bf u}$, $E_f$ is the
kinetic energy restricted to the forcing band,
$E_f = \frac{1}{2} \sum_{0< |{\bf k} | < k_f} | \tilde{{\bf u}} ({\bf k}, t)|^2$,
and one can view the external forcing as an energy pumping through the low wave numbers with
the energy pumping rate $\e_f$, from the following energy equation
\begin{equation}
\frac{d}{dt} \sum_{{\bf k}} |\tilde{{\bf u}}|^2 = -\nu \sum_{{\bf k}} |{\bf k} \tilde{{\bf u}}|^2 + \e_f ,
\end{equation}
here the first term on the right hand side is the turbulence energy dissipation rate. Through
such energy pumping that balances the energy dissipation, one can drive the turbulence into a statistically steady state of homogeneous isotropic turbulence. The external forcing here has
been well tested by other researchers \cite{KI06} \cite{LM15}. For our numerical simulations here,
we choose $k_f = 2.5$ and $\e_f = 0.1$. The Reynolds number is specifically defined by
$Re = \frac{VL}{\nu}$, where $V$ is the rms velocity, and $L$ is the integral length scale
$L = \frac{3\pi}{8E} \sum_{{\bf k}} \frac{| \tilde{{\bf u}} ({\bf k} )|^2}{|{\bf k} |}$, where $E$ is the kinetic energy. The large eddy turnover time is given by $T_0 = L/V$. In our simulations, the large eddy turnover time is around $2$. All our direct numerical simulations are
well resolved to scales smaller than the Kolmogorov scale. We also test resolution to scales
much smaller than the Kolmogorov scale to make sure our resolution is statistically sufficient.

\begin{figure}[ht]
\centering
\includegraphics[width=3.25in,height=2.25in]{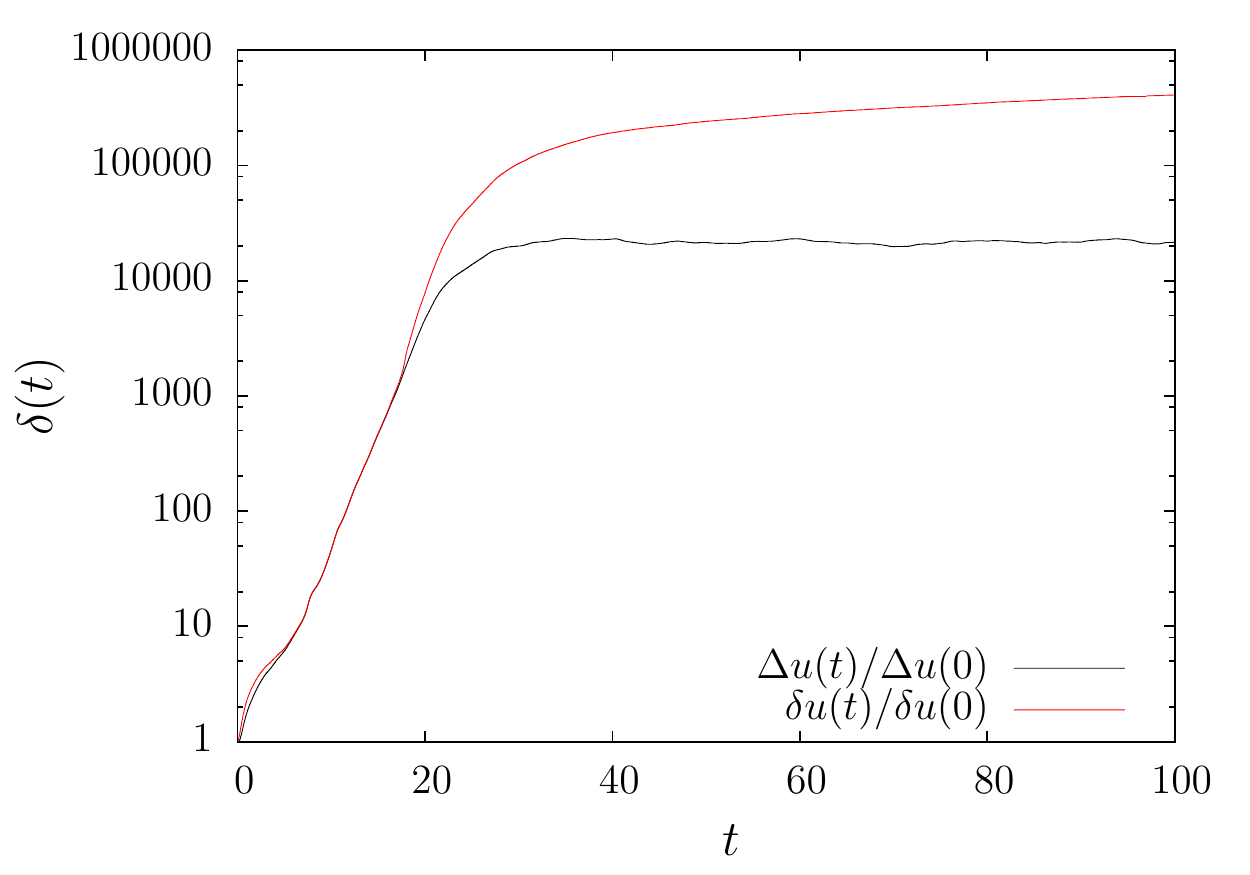}
\caption{The divergence of the linear amplification and the nonlinear amplification of a
perturbation with initial energy norm $\Delta u (0) = \delta u (0) = 0.01$ (\ref{nan})-(\ref{lan})
and the Reynolds
number $Re = 130$. $\delta (t)$ represents either $\frac{\Delta u (t)}{\Delta u (0)}$ or
$\frac{\delta u (t)}{\delta u (0)}$. Divergence and nonlinear saturation happens around $t=20$
which is about $10$ large eddy turnover time $10 T_0$.}
\label{div}
\end{figure}

After we run our direct numerical simulation to a statistically steady state of homogeneous isotropic
turbulence, we introduce a perturbation by first making a copy of the velocity field at a time step and then not calling the external forcing at that time step for one copy of the velocity field. Thus
the perturbation will be in the band of wave numbers of the external forcing (\ref{EF}), 
$0< |{\bf k} | < k_f$. After that time step, both fields will call the external forcing normally. 
We can then track the nonlinear or linear evolution of the perturbation as described before.
We denote the nonlinear evolution of the perturbation by $\Delta {\bf u} ({\bf x} , t)$, the linear
evolution of the perturbation by $\delta {\bf u} ({\bf x} , t)$, and their energy norms by $\Delta (t) = \Delta u (t)$
and $\delta u (t)$:
\begin{equation}
\Delta (t) = \Delta u (t) = \left (\sum_{{\bf k}} |\widetilde{\Delta {\bf u}} ({\bf k} , t)|^2 \right )^{1/2} ,
\label{nan}
\end{equation}
\begin{equation}
\delta u (t) = \left (\sum_{{\bf k}} |\widetilde{\delta {\bf u}} ({\bf k} , t)|^2 \right )^{1/2} .
\label{lan}
\end{equation}
We notice that the nonlinear amplification $\Delta u (t)$ is bounded in time (see the Supplementary Material of \cite{BH18}), while the linear amplification $\delta u (t)$ is unbounded in time. Our
prediction (\ref{OP}) is for the linear amplification $\delta u (t)$. During the early stage (small
time) amplification, the linear amplification $\delta u (t)$ is a good approximation of the
nonlinear amplification $\Delta u (t)$, and our prediction (\ref{OP}) also applies to the nonlinear amplification $\Delta u (t)$. Then the linear amplification and the nonlinear amplification
are going to diverge, that is the transition to nonlinear saturation, and the linear amplification
is not a good approximation of the nonlinear amplification any more. We want to emphasize that
the nonlinear amplification is what is happening in reality. We are therefore most interested in
the initial stage amplification, before the nonlinear saturation. Of course, the nonlinear saturation depends on the initial energy norm of the perturbation $\Delta u (0)$, but it also depends on the Reynolds number. In Figure \ref{div}, we show the nonlinear saturation for an initial perturbation
with the energy norm $\Delta u (0) = \delta u (0) = 0.01$ and the Reynolds number $Re = 130$.
Divergence of the linear and nonlinear amplifications happens around $t=20$
which is about $10$ large eddy turnover time $10 T_0$. Increasing the Reynolds number will
drastically reduce the saturation time. Increasing the energy norm of the initial perturbation
of course will also reduce the saturation time. From now on, we will mainly focus on the time before
nonlinear saturation to verify our prediction (\ref{OP}).

\begin{figure}[ht]
\centering
\includegraphics[width=3.25in,height=2.25in]{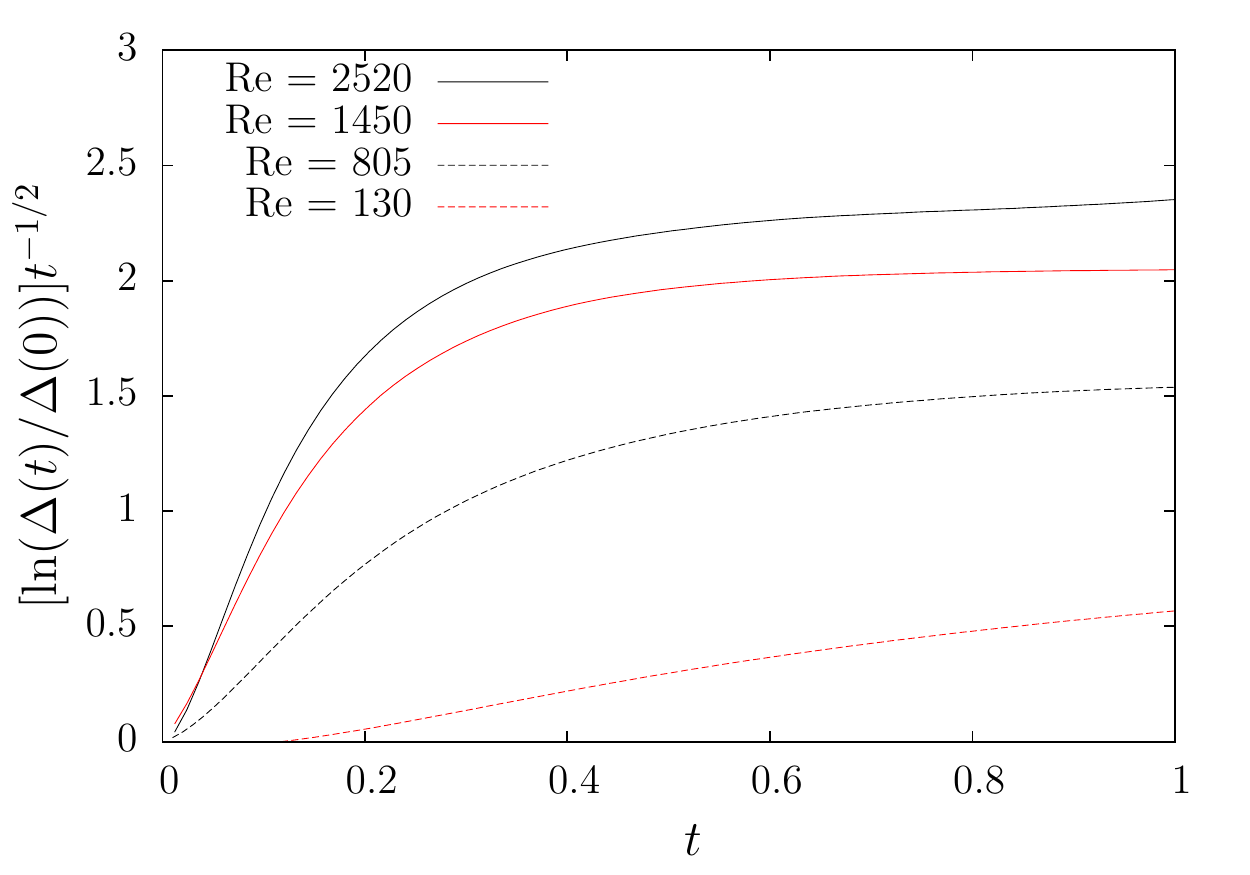}
\caption{The nonlinear amplifications of perturbations for different Reynolds
numbers $Re = 130, 805, 1450, 2520$. $\Delta (t)$ is the energy norm defined in (\ref{nan}). The
horizontal portions of the curves represent amplifications of $e^{c \sqrt{t}}$ in $t$ as we predicted in (\ref{OP}).}
\label{amp}
\end{figure}

In Figure \ref{amp}, we show the nonlinear amplifications
of perturbations for the Reynolds numbers $Re = 130, 805, 1450, 2520$. Except in $t \ra 0^+$ limit,
we clearly observed the amplifications of $e^{c \sqrt{t}}$ in $t$ as we predicted in (\ref{OP}).
We use $c$ to represents a generic constant in this paper.
In the $t \ra 0^+$ limit, numerical resolution can never be sufficient and generates false results
as shown by an explicit example in \cite{FL18}. Our extensive numerical simulations here and earlier
show that the amplifications of $e^{c \sqrt{t}}$ in $t$ are generically observed \cite{FL18}.

\begin{figure}[ht]
\centering
\includegraphics[width=3.25in,height=2.25in]{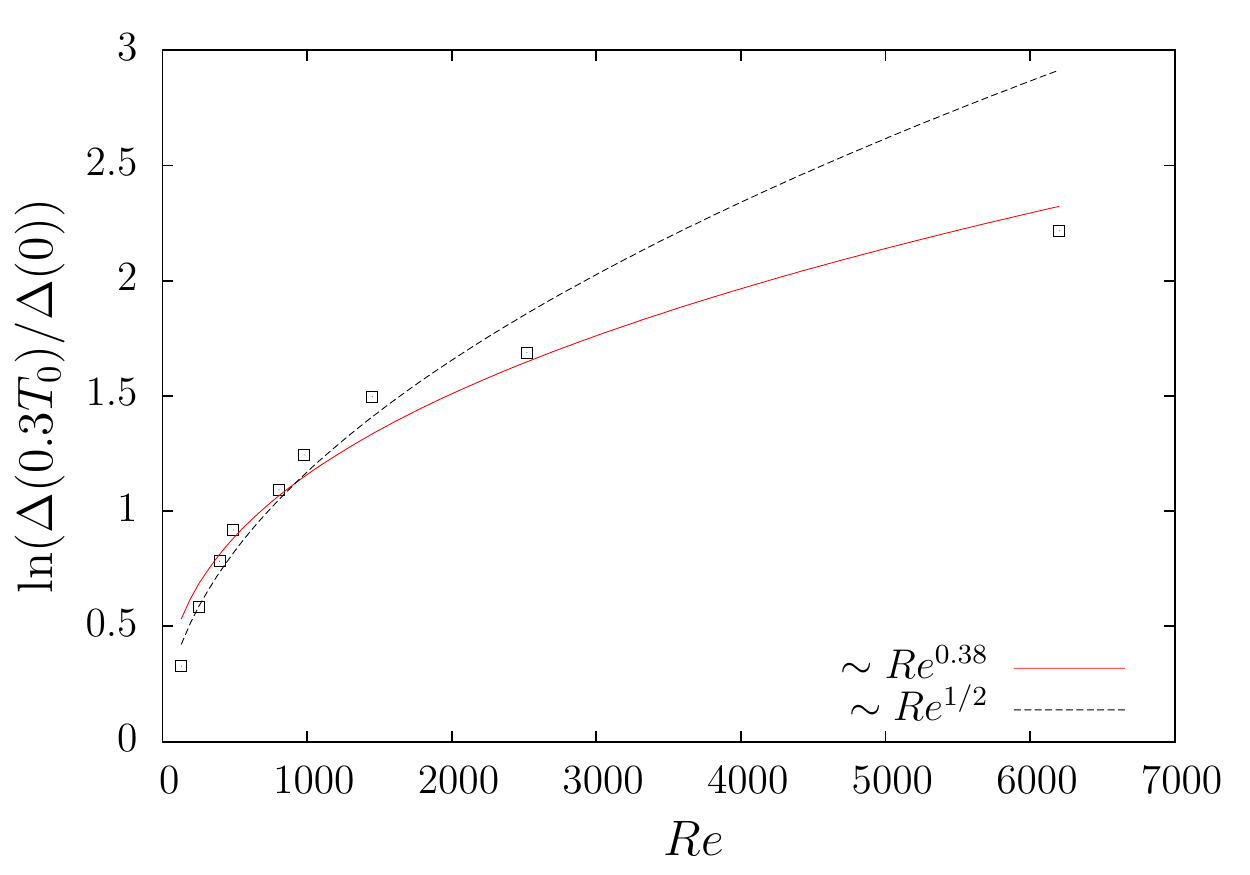}
\caption{The Reynolds number dependence of the nonlinear amplifications of perturbations
for different Reynolds up to the time $t = 0.3 T_0$, where $T_0$ is the large eddy turnover time
which is around $2$. The dashed curve is a fit to $\sim \sqrt{Re}$, and the red (grey) curve
is a fit to $\sim Re^{0.38}$.}
\label{ARe}
\end{figure}

In Figure \ref{ARe}, we show the dependence of the nonlinear amplifications of perturbations up
to the time $t = 0.3 T_0$ on the Reynolds number, where $T_0$ is the large eddy turnover time
which is around $2$. The data fit well with $e^{c \sqrt{Re}}$ as we predicted in (\ref{OP}).
Thus both the $\sqrt{t}$ and the $\sqrt{Re}$ features in the exponent of our prediction (\ref{OP})
are verified here in the fully developed homogeneous isotropic turbulence regime. 

In conclusion, built upon our earlier low resolution numerical verifications of our prediction
(\ref{OP}) on the maximal amplification of perturbations in fully developed turbulence \cite{FL18},
here we conduct large direct numerical simulations with sufficient resolutions on fully developed homogeneous isotropic turbulence, and we have verified our prediction (\ref{OP}). In particular,
our numerical simulations show that the amplification of perturbations behaves as $e^{c\sqrt{t}}$
in $t$ and $e^{c\sqrt{Re}}$ in $Re$ as we predicted in (\ref{OP}). Thus amplifications of
perturbations in fully developed turbulence are much faster than exponential in contrast to
the exponential amplifications of perturbations in chaos. Turbulence should appear more violent
than chaos. Such superfast amplifications of perturbations naturally lead to superfast nonlinear
saturations, and we have demonstrated such nonlinear saturations. We conclude that fully developed
turbulence is generated, developed and maintained by such constant superfast amplifications of ever existing perturbations. 
We believe that this theory better explains what is observed in fully developed turbulence than the chaos theory.

We believe that our discovery here will better guide turbulence engineering, design and prediction such as the 
Ensemble Weather Forecasting in Meteorology.

\begin{acknowledgments}
This work has used resources from the Edinburgh Compute and Data Facility (http://www.ecdf.ed.ac.uk) and
ARCHER (http://www.archer.ac.uk). A.B acknowledges support from the UK Science and Technology 
Facilities Council whilst R.D.J.G.H is supported by the
UK Engineering and Physical Sciences Research Council (EP/M506515/1).
\end{acknowledgments}

\end{document}